\documentclass[aps,epsf,preprint]{revtex4}

\def\1{{\bf 1}}
\def\[{\left[}
\def\]{\right]}
\def\be{\begin{eqnarray}}
\def\ee{\end{eqnarray}}
\def\nn{\nonumber}
\def\({\left(}
\def\){\right)}
\def\bk#1{\langle#1\rangle}
\def\eq#1{(\ref{#1})}

\def\a{\alpha}
\def\r{\rho}
\def\o{\omega}

\def\f{\phi}
\def\q{\psi}
\def\G{{\cal G}}

\def\x{\times}
\def\ket#1{|#1\rangle}

\def\vss{\vskip.1cm}
\def\p{\partial}

\def\t{\tau}
\def\s{\sigma}

\def\labels#1{\label{#1}}
\def\edc{\end{document}}
\def\P{{\cal P}}

\def\bn{\begin{enumerate}}
\def\i{\item}
\def\en{\end{enumerate}}
\def\b{\beta}
\def\g{\gamma}
\def\ol{\overline}
\def\rd{\sqrt{2}}
\def\rt{\sqrt{3}}

\def\rs{\sqrt{6}}
\def\diag{{\rm diag}}

\def\th{\theta}
\def\ba{\begin{array}}
\def\ea{\end{array}}
\def\bc{\begin{center}}
\def\ec{\end{center}}

\def\Tr{{\rm Tr}}
\def\ul{\underline}

\usepackage{graphicx}
\graphicspath{%
    {converted_graphics/}
    {/}
}
\begin{document}

\title{A Horizontal Symmetry for Leptons and Quarks}
\author{C.S. Lam}
\address{Department of Physics, McGill University\\
 Montreal, Q.C., Canada H3A 2T8\\
and\\
Department of Physics and Astronomy, University of British Columbia,  Vancouver, BC, Canada V6T 1Z1 \\
Email: Lam@physics.mcgill.ca}

\begin{abstract}
A generic valon potential invariant under $U(1)\x SO(3)$ is used to determine whether the horizontal symmetry of leptons is
$A_4$, $S_4$, or some other finite subgroups of $SO(3)$. Valons in the potential
are assigned  $SO(3)$ spins  0, 1, and 2, for these are the only ones
 that can couple to the three generation of fermions with horizontal spin 1. This potential causes a
 breakdown into three possible phases with three 
 different symmetries. Phase I
has an $A_4$ symmetry suitable for leptonic mixing. Phase II has an $SO(2)$ symmetry and phase III has a $Z_2\x Z_2$
symmetry, both  capable of describing Cabibbo mixing of quarks. Phase II has to be rejected on phenomenological ground,
but phase III yields block diagonal and hierarchical mass matrices appropriate for quarks. No other 
non-abelian symmetry such as $S_4$ is present.

\end{abstract}
\narrowtext
\maketitle

\section{Introduction}
Symmetry is the foundation of particle physics, so 
a new symmetry might provide the right key to
 a door beyond the Standard Model. One candidate is a symmetry of the three fermion generations which interact
identically under the Standard Model.  Such a (horizontal) symmetry is much discussed, but its existence is not confirmed. 
Owing to the large mass difference and 
 mixing, if it exists it has to be badly broken. That makes it difficult to be certain of the existence of a
symmetry, much less what it is. While mass difference makes it difficult to locate the symmetry, amusingly
mixing pattern does help to find it \cite{LAM}.

The hunt for  horizontal symmetry received an important stimulus when neutrino mixing was found to be so regular
that it could be described  by
a tri-bimaximal mixing matrix (TBM) \cite{HPS}. It was also found that this pattern of mixing could be explained by models with
 an $A_4$ horizontal symmetry \cite{A4}.
$A_4$ is the smallest non-Abelian group that has a three-dimensional irreducible representation, making it attractive
because most of the existing symmetries are in some sense described by the smallest possible groups.

Nevertheless, the natural symmetry associated with TBM is the group $S_4$ \cite{LAM}, not its subgroup $A_4$. The reason why $A_4$ works in those models can be traced back to 
a dynamical choice,  allowing an accidental symmetry to promote $A_4$ to $S_4$. From the symmetry point of view,
$S_4$ is a more natural group, but it is not the smallest group that $A_4$ is. 
Is the leptonic symmetry $A_4$, or $S_4$, or something else? One possible way to decide
is to appeal to dynamics.

Both $A_4$ and $S_4$ are finite subgroups of $SO(3)$, so if
 the original unbroken horizontal symmetry is $SO(3)$, dynamics of its Higgs valons might determine whether $SO(3)$ breaks
down to $A_4$, or $S_4$, or some other finite subgroup, thereby telling us what the preferred symmetry of leptons is.
 It is this possibility that we want to investigate in the present article.

We shall refer
to the quantum number of this horizontal $SO(3)$ as {\it horizontal spin}, but the word `horizontal'  is often omitted for simplicity
because we do
not discuss any other kind of spin or symmetry here.
With three generations of leptons, it is natural to assign them a horizontal spin 1. Valons 
that couple to them in  Yukawa terms must then have horizontal spins 0, 1, and 2. If
$SO(3)$ indeed breaks down to $S_4$ or $A_4$ in the lepton sector, with the proper vacuum alignments
neutrino mixing is automatically TBM, then the three
Yukawa coupling constants can be used to fit the three fermion masses. 

Unfortunately it is known that whatever the dynamics is, 
it is impossible for $SO(3)$ to be spontaneously broken down to either $A_4$ or $S_4$, as long as only valons of horizontal spins 0, 1, and 2 are involved \cite{SO3}. The reason will be reviewed in more detail in the 
next section, but 
essentially it is because valons of spins 1 and 2 cannot contain a state invariant under either $A_4$ or $S_4$. 
However, in the case of $A_4$, there is a spin-2 valon state which 
differs from an invariant state under an $A_4$ transformation only by a phase factor. No such state exists for $S_4$.
It is therefore conceivable that in a theory invariant under $U(1)\x SO(3)$, the extra $U(1)$ may provide a compensating phase factor
to render this state invariant under $A_4$, though such an invariant state is still not present in $S_4$. This is indeed the case
as we will show later, so in this way
the dynamics of $U(1)\x SO(3)$ chooses $A_4$ to be the appropriate leptonic symmetry group over $S_4$. The basic reason for that to be so
is because $A_4$ is a smaller group, so a state invariant under the subgroup $A_4$ may not be invariant under the full group $S_4$.

We shall refer to the quantum number of this $U(1)$ as {\it horizontal charge}.

The extra symmetry $U(1)$ can be understood as providing a second protection for fermion masses. Isospin and hypercharge provides
one protection because the left-handed and right-handed leptons have different hypercharge and isospin, preventing a mass term to be formed.
Similarly, the valons here possess a non-trivial horizontal charge, forcing the horizontal charges of the
 left-handed and the right-handed fermions
to be different, thereby also forbidding a mass term to be formed. In this way a second protection of
the fermion masses is obtained.

A serious impediment to the establishment of a credible
 horizontal symmetry is the qualitative difference between quark  and lepton mixings. Instead of having
 two large mixing angles as is the case for neutrinos, quark mixings are all small. The regularity of quark mixing is
 not given by anything resembling a TBM pattern, but merely by a hierarchical structure.
Quark mixing is dominated by the mixing of the first two generations, 
with the mixing of the second and the third generations quite a bit smaller,
and that between the first and the third generations the smallest. 
In the approximation of neglecting
the smaller mixings with the third generation, the mixing matrix is block diagonal, with the third generation unmixed.
This means that the mass matrices of the up- and down-quarks can be simultaneously
put into a block-diagonal form $\diag(m_{12}, m_3)$, where $m_3$ is the mass of the third generation, and $m_{12}$
is an arbitrary 2$\x 2$ matrix from which Cabibbo mixing of the light masses are determined. We shall refer to 
a block-diagonal mass matrix of this form as {\it hierarchical}
if $|m_3|$ is much larger than the matrix elements of $m_{12}$. From the mass spectra of quarks, the mass matrices of the quarks
seem to be not only block-diagonal approximately, but also hierarchical.

If horizontal symmetry is really fundamental, how come lepton mixing is governed by a symmetry such as $A_4$,
but quark mixing relies only on a hierarchical structure without any apparent symmetry? 
Is it possible that this qualitative difference
is just an illusion at low energy, with both of them descending from a common symmetry at high energy? 

Before proceeding along this line it should be mentioned that models can be constructed 
in which both quark and lepton mixings are based on $A_4$, but at the expense of
giving up exact TBM. We will not pursue that possibility here.

Interestingly 
the group $U(1)\x SO(3)$ used to decide the leptonic symmetry to be $A_4$ could also serve as that common symmetry at high energy.
This is because the group breaks spontaneously into three possible phases with different symmetries, some suitable for leptons, 
and others suitable for quarks. Phase I has a tetrahedral symmetry $A_4$
that can be employed to explain leptonic mixing, and
Phases II and III can produce hierarchical block-diagonal mass matrices that are appropriate for
quark mixing. However, Phase II fails because the mixing is between a heavy and a light quark,  but Phase III is perfectly viable.

The details of the $U(1)\x SO(3)$ dynamics and the symmetry of the three phases will be dealt with in Sec.~III. 
The fermion mass and mixing matrices in  different phases will be discussed in Sec.~IV. There are also two appendices
containing mathematical details too cumbersome to be discussed in the main text, and a concluding section V.

\section{Horizontal Spin}
Suppose $SO(3)$ is a horizontal symmetry at some high energy scale. It was mentioned in the Introduction that no valon potential
can break it down spontaneously to $A_4$ or $S_4$ if the valons have a horizontal spin less than 3 \cite{SO3}. The reason 
for that is reviewed in this section.

If $SO(3)$ can be spontaneously broken down to a subgroup $\G$, then the valon must contain a state invariant under $\G$. Using the known
characters of $SO(3)$ and $\G$, we can work out what irreducible representations (IR) of $\G$ that a horizontal spin-$J$ valon contains. The
result is shown in Table 1 for $J=0, 1, 2, 3$, and $\G=A_4, S_4, A_5$.

$$\ba{|c|cccc|ccccc|ccccc|}\hline
\G&&&A_4&&&&S_4&&&&&A_5&&\cr\hline
{\rm IR}&1&1'&1''&3&1&1'&2&3'&3&1&4&5&3'&3\cr\hline\hline
J=0&1&0&0&0&1&0&0&0&0&1&0&0&0&0\cr\hline
J=1&0&0&0&1&0&0&0&0&1&0&0&0&0&1\cr\hline
J=2&0&1&1&1&0&0&1&1&0&0&0&1&0&0\cr\hline
J=3&1&0&0&2&0&1&0&1&1&0&1&0&1&0\cr\hline
\ea$$
\bc Table 1. Multiplicity of horizontal spin-$J$ state in irreducible representations (IR) of $\G$\ec

Let us see what the table tells us.
\bn
\i Other than the trivial $J=0$ state, the only state invariant under $\G$ (transforms like {1}) is
the $J=3$ state, with $\G=A_4$. There are no invariant states for $J=1$ or $J=2$.
\i Although there is no invariant state for $J=2$ and $\G=A_4$, two of the states ($1', 1''$) are pseudo-invariant,
in that under an $A_4$ transformation, these states only gain a phase. There is no state in $\G=S_4$ nor $\G=A_5$
which has that property for $J\le 2$.
\i This observation gives us hope that if we enlarge $SO(3)$ to $U(1)\x SO(3)$,  an additional phase 
from $U(1)$ can be incorporated to cancel the phase of the pseudo-invariant states, thereby turning
them into invariant states of $A_4$. If that is indeed the case, then
$A_4$ could be a descendent of $U(1)\x SO(3)$, though not $S_4$, nor $A_5$. 

\en

\section{${\bf U(1)\x SO(3)}$ Higgs Dynamics}
It will be shown in this section that a generic $U(1)\x SO(3)$ potential of valons with horizontal spins 0 and 2 has three solutions.
These solutions differ from one another in order parameters and in symmetry.

Let $\f=(a_2,a_1,a_0,a_{-1},a_{-2})$ be a complex valon field with horizontal spin 2, and $b$  a complex valon field with
horizontal spin 0.  The subscript of $a$ indicates the horizontal
angular momentum along some fixed $z$-axis to be defined later. Let $u(\xi)$ be a member of the $U(1)$ group obeying $u(\xi_1)u(\xi_2)=u(\xi_1
+\xi_2)$, and let $u(\xi)\f=e^{i\xi}\f,\ u(\xi)\f^*=e^{-i\xi}\f^*,\ u(\xi)b=e^{i\xi}b,\ u(\xi)b^*=e^{-i\xi}b^*$.
In other words, in additional to the horizontal spin, we also 
assign $\f$ and $b$ a {\it `horizontal charge'} $+1$ and $\f^*$ and $b^*$ a horizontal charge $-1$.

A real renormalizable 
Higgs potential of $\f, \f^*, b,$ and $b^*$
invariant under $U(1)\x SO(3)$ may contain only $\f^*\f,\ b^*b,\  \f^*\f^*\f\f,\ b^*b^*bb,\ b^*b^*\f\f,\ \f^*\f^*bb $ 
types of terms. If we choose to  couple
first $\f\f$, and separately $\f^*\f^*$, before putting them together in the quartic terms, 
then the most general Higgs potential can be seen to be
\be
V&=&V_\f+V_b+V_{\f b},\nn\\
V_\f&=&\bk{\f\f|g_0\P_0+g_2\P_2+g_4\P_4|\f\f}
-\mu^2\bk{\f|\f},\nn\\
V_b&=&\r (b^*b)^2-\nu^2(b^*b),\nn\\
V_{\f b}&=&\tau\bk{\f|\f}b^*b+\s^* b^{*2}\th+\s b^2\th^*,
\labels{V}\ee
where $\P_J$ is the operator projecting into states with horizontal spin $J$. 
 The reality of $V$ implies that the coupling constants 
$g_J, \r, \t, \mu^2, \nu^2$ are real. The quantity 
\be
\th=\sum_{m=-2}^2(-)^ma_ma_{-m}\labels{theta}\ee
is proportional to the component of the compound state $\ket{\f\f}$ with zero horizontal angular momentum. See \eq{ketc} below.
The potential $V_\f$, without the $\mu^2$ term,
 has been used to study $d$-wave superconductivity, and Bose-Einstein condensates
of spin-2 cold atoms \cite{COLD}.

$V_\f$ in \eq{V} can be expressed in a more explicit form by getting rid of the projection operators. If $\vec J_1$ and $\vec J_2$ are the
horizontal angular momentum operators of valons 1 and 2, then one relation between the projection operators can be derived by considering
\be\bk{\f\f|(\vec J_1+\vec J_2)^2|\f\f}=\bk{\f\f|0\P_0+6\P_2+20\P_4|\f\f}=12+2\bk{\f|\vec J_1|\f}_1\cdot\bk{\f|\vec J_2|\f}_2.\labels{p1}\ee
Choosing the horizontal-spin condensate $\bk{\f|\vec J|\f}$  along the $z$-axis, we can write
$\bk{\f|\vec J_1|\f}_1\cdot\bk{\f|\vec J_2|\f}_2=\bk{\f|J_z|\f}^2:=\bk{J_z}^2$. Together with
 the completeness relation $\P_0+\P_2+\P_4={\bf 1}$, we can solve $\P_2$ and $\P_4$ in terms of $\P_0$ to get
\be\P_4={1\over 7}(3+\bk{J_z}^2+3\P_0),\quad \P_2={1\over 7}(4-\bk{J_z}^2-10\P_0).\labels{p24}\ee
Substituting \eq{p24} into \eq{V}, and writing 
$5\bk{\f\f|\P_0 |\f\f}=|\th|^2$,
the potential becomes
\be V_\f=\alpha K^2+\beta\bk{J_z}^2+\gamma|\theta|^2-\mu^2K,\labels{V2}\ee
where
\be \alpha={1\over 7}(3g_4+4g_2),\quad \beta={1\over 7}(g_4-g_2),\quad \gamma={1\over 5}(g_0-g_4)+{2\over 7}(g_4-g_2),\labels{abc}\ee
and
\be 
K=\bk{\f|\f}=\sum_m|a_m|^2,\quad \bk{J_z}=\sum m|a_m|^2,\quad
\th=\sum(-)^ma_{-m}a_m. \labels{ketc}\ee

No attempt has been made to include couplings with a spin-1 valon in the potential to keep it within
manageable complexity. We will simply assume 
that the presence of a spin-1 valon does not affect
the main conclusions reached without it. This would be the case for example if the coupling with it is  weak,
or that its natural energy scale is much lower than that of the spin-2 or the spin-0 valons. Moreover, as we will see later,
the phase diagram obtained from spin-2 valons is not altered by its coupling with a spin-0 valon, so maybe the same thing is true
for a coupling with a spin-1 valon as well.

\subsection{Equations of Motion}
There are 12 equations of motion, $\p V/\p a_m^*=0=\p V/\p a_m$ ($m=2,1,0,-1,-2$), and $\p V/\p b^*=0=\p V/\p b$, consisting
 of the following six  and their complex conjugates:
\be
&\alpha_0 a_m+\beta\bk{J_z}ma_m+\gamma(-)^m\th a^*_{-m}+{1\over 2}\s b^2(-)^ma^*_{-m}=0&,\labels{eqm}\\
& \a_0:=\a-\mu^2/2K+\t|b|^2/2K&;\labels{alpha0}\\
&\r_0|b|^2b+\s^*\th b^*=0&,\labels{eqmb}\\
&\r_0:=\r-\nu^2/2|b|^2+\t K/2|b|^2&.\labels{rho0}
\ee

The cubic equations  
in $a_m, a_m^*, b$ and $b^*$ are more easily solved by first turning them into quadratic equations of 
the {\it order parameters} $K, \bk{J_z},
\th, \th^*$, and $|b|^2$. This can be achieved by multiplying \eq{eqm}  by $a_m^*$, and separately by $(-)^ma_{-m}$, then summing
 over all $m$.
In this way we get
\be
&\alpha_0K^2+\beta\bk{J_z}^2+\gamma|\th|^2+{1\over 2}\s b^2\th^*=0,&\qquad\qquad{\rm and} \labels{em1}\\
&[(\alpha_0 +\gamma)\th+{1\over 2}\s b^2]K=0.&\labels{em2}\ee
It follows that $\s b^2\th^*$ must be real.

Similarly, by multiplying \eq{eqmb} by $b^*$, we get
\be
\r_0|b|^4+\s^*\th b^{*2}=0.\labels{b4}\ee

Substituting \eq{em1} and \eq{b4} into \eq{V2} and \eq{V}, a simplified expression for the Higgs energy emerges:
\be
V=-{1\over 2}(\mu^2K+\nu^2|b|^2)+{1\over 4}(\s b^2\th^*+\s^*b^{*2}\th).\labels{V3}\ee

\subsection{Solutions}
Solutions can be classified by the order parameters $\bk{J_z}, \th, K$, and $|b|^2$. First, assume $b\not=0$.

\vskip.5cm
\noindent$\fbox{$b\not=0$}$
\vskip.3cm

Consider separately the cases $\th\not=0$ and $\th=0$.
If $\th=0$, then \eq{b4} implies $\r_0=0$ and \eq{em2} implies $\s=0$. Using \eq{rho0}, we obtain
\be 2\r |b|^2=\nu^2-\t K.\labels{A0}\ee
Now consider one after another the two cases, $\bk{J_z}=0$  and $\bk{J_z}\not=0$.	
\bn 
\i[{\bf I}.]\ $\th=0, \bk{J_z}=0$.\quad From \eq{eqm}, $\a_0=0$, hence from \eq{alpha0},
\be 2\alpha K=\mu^2-\t |b|^2.\labels{A1}\ee
\eq{A0} and \eq{A1} can be used to solve $|b|^2$ and $K$ to get
\be |b|^2={-\t \mu^2+2\a\nu^2\over -\t^2+4\a\r}:=|b_I|^2,\quad K={2\r\mu^2-\t\nu^2\over -\t^2+4\a\r}:=K_I.\labels{A2}\ee
Higgs energy can then be obtained from \eq{V3} to be
\be V=V_I:=-{\r\mu^4+\a\nu^4-\tau\mu^2\nu^2\over -\t^2+4\a\r}.\labels{A3}\ee

One solution for $\f$ that gives rise to $\th=\bk{J_z}=0$ is
\be \f_I=\sqrt{K\over 3}\(1,0,0,\rd,0\).\labels{psi0}\ee
It will be shown in Appendix A that this solution is unique up to a $U(1)\x SO(3)$ transformation.

\i[{\bf II}.]\  $\th=0, \bk{J_z}\not=0$.\quad
\eq{eqm} implies one and only one $a_m$ is non-zero, and that this cannot be $a_0$ for otherwise $\bk{J_z}=0$.
It is easy to see that $\bk{J_z}=mK$, so according to \eq{eqm} and \eq{alpha0}, $\a_0+\b m^2=0$, hence
\be
2(\a+\b m^2)K=\mu^2-\t |b|^2.\labels{A4}\ee
$K$ and $|b|^2$ can now be solved from \eq{A0} and \eq{A4} to give
\be |b|^2={-\t \mu^2+2(\a+m^2\b)\nu^2\over -\t^2+4(\a+m^2\b)\r}:=|b_{II}|^2,\quad 
K={2\r\mu^2-\t\nu^2\over -\t^2+4(\a+m^2\b)\r}:=K_{II}.\labels{A5}\ee
 
Higgs energy is then obtained from \eq{V3} to be
\be V=V_{II}:=-{\r\mu^4+(\a+m^2\b)\nu^4-\t\mu^2\nu^2\over -\t^2+4(\a+m^2\b)\r}.\labels{A6}\ee
Note that \eq{A1} to \eq{A3} can be obtained from \eq{A4} to \eq{A6} simply by setting $m=0$. The alignment of the spin-2 valon is
\be \f_{II}=\sqrt{K}(1,0,0,0,0),\quad \f'_{II}=\sqrt{K}(0,1,0,0,0),\labels{A11}\ee
for $m=2$ and $m=1$ respectively. The solutions with $m=-2$ and $m=-1$  can be obtained from these by a $\pi$-rotation about the $y$-axis.

\i[{\bf III}.]\ $\th\not=0$.\quad In this case we must have $\bk{J_z}=0$.  Otherwise, as argued before, there is one and only one $a_m\not=0$, 
and this is not $a_0$. As a result, according to \eq{theta} or \eq{ketc}, we must have $\th=0$, contradicting the 
assumption that $\th\not=0$.

The solution in this case is considerably more complicated because we are no longer forced
to have $\s=0$.

Multiply \eq{em2} by $\th^*$, then compare it with \eq{em1} and with \eq{b4}, we get
\be
\a_0K^2+\g|\th|^2&=&\r_0|b|^4/2,\labels{emx}\\
(\a_0+\g)|\th|^2&=&\r_0|b|^4/2,\labels{emy}\ee
which shows that $|\th|^2=K^2$. It also shows that $\r_0$ and $\a_0+\g$ are both positive, or both negative.

Let us first assume both of them to be negative. Then taking the absolute value of \eq{em2} and \eq{eqmb}, and use \eq{alpha0}
and \eq{rho0}, we get
\be (\a_0+\g)K=(\a+\g)K+{1\over 2}(\t |b|^2-\mu^2) &=&-{1\over 2}|\s||b|^2,\labels{A7}\\
 \rho_0|b|^2=\r |b|^2+{1\over 2}(\t K-\nu^2)&=&-|\s|K.\labels{A8}\ee
These two yield the solution $|b|^2=|b_{III}|^2$ and $K=K_{III}$, with
\be
|b_{III}|^2={2(\a+\g)\nu^2-(2|\s|+\t)\mu^2\over 4(\a+\g)\r-(\t+|\s|)(\t+2|\s|)},\quad 
K_{III}={-(\t+|\s|)\nu^2+2\r\mu^2\over 4(\a+\g)\r-(\t+|\s|)(\t+2|\s|)}.\labels{A9}\ee
They cause the Higgs energy in \eq{V3} to be
\be
V&:=&V_{III}=-{\mu^4N_1+\mu^2\nu^2N_2+\nu^4N_3\over \[4(\a+\g)\r-(\t+|\s|)(\t+2|\s|)\]^2},\nn\\
N_1&=&\r\[4\r(\a+\gamma)-\t(\t+2|\s|)\],\nn\\
N_2&=&(\t+2|\s|)\[-4\r(a+\gamma)+(\t+|\s|)^2\],\nn\\
N_3&=&\[-(\t+|\s|)^2(\a+\gamma)+4\r(\a+\gamma)^2\].
\labels{A10}\ee

If both $\r_0$ and $\a_0+\g$ are positive, then the right-hand side of equations \eq{A7} and \eq{A8} change a sign. In that case,
solutions \eq{A9} and \eq{A10} are still valid provided we change $|\s|$ in every expression to $-|\sigma|$.

One spin-2 vacuum alignment that gives rise to $|\th|=K$ and $\bk{J_z}=0$ is
\be \f_{III}=\sqrt{K}(\sin\eta/\rd,0,\cos\eta ,0,\sin\eta /\rd),\labels{nematic}\ee
	where $\eta$ is an arbitrary angle between 0 and $\pi$. It will be shown in Appendix A that all other solutions
can be obtained from this one by a $U(1)\x SO(3)$ transformation.

\en
For convenience, these three type of solutions with $b\not=0$ are summarized in Table 2. The last column indicates
the condition under which the solution is valid, and the numbers within parentheses are equation numbers describing
the solutions.

$$\ba{|c|c|c|c|c|c|c|c|}\hline
{\rm sol}&|\th|&\bk{J_z}&|b|^2&K&V&\f&{\rm cond}\cr\hline
{\bf I}&0&0&\eq{A2}&\eq{A2}&\eq{A3}&\eq{psi0}&\s=0\cr\hline
{\bf II}&0&mK&\eq{A5}&\eq{A5}&\eq{A6}&\eq{A11}&\s=0\cr\hline
{\bf III}&K&0&\eq{A9}&\eq{A9}&\eq{A10}&\eq{nematic}&\cr
\hline\ea$$
\bc Table 2. A summary of the $b\not=0$ solutions\ec 
\vskip1cm
\noindent$\fbox{$b=0$}$
\vskip.5cm
It is clear from \eq{eqmb} that there are solutions with $b=0$. It is also clear from \eq{eqm}
that the solutions for spin-2 valons are identical to those obtained above, after setting
$\s=\t=0$. From \eq{V3}, the potential $V$ is also identical to those obtained before by
setting $\s=\t=\nu^2=0$.

\subsection{Energy and Phases}
Stability demands a condensate to have the lowest energy. It is therefore necessary to determine which
of the solutions in Table 2 fits that role. 
The result turns out to depend only on the relative sizes of the parameters $\a, \b$, and $\g$, as shown
in the phase diagram in Fig.~1, and not on the parameter $\t$.  Here is how that conclusion is arrived at.

First of all, if $\s\not=0$, then the only solution and the only phase is {III}, so there is no phase diagram
and no need for a further discussion.
From now on, we will assume $\s=0$.

The parameters $\mu$ and $\nu$ have the dimension of energy, and the rest 
are dimensionless. In the absence of couplings between the (horizontal) spin-2 and spin-0
valons ($\s=\t=0$), they define the energy scales of the spin-2 condensate $K$ and the spin-0 condensate $|b|^2$, respectively.
The spin-0 condensate is given by $|b|^2=\nu^2/2\r$, and the spin-2 condensate is given by either
{I}. $K=\mu^2/2\a$, {II}. $K=\mu^2/2(\a+m^2\b)$, or {III}. $K=\mu^2/2(\a+\g)$, depending on which
of the solution has the least energy.  Note that since $|b|^2$ and $K$ are non-negative,
the parameters $\rho, \a, (\a+\g), (\a+4\b)$ must also be non-negative.

The energy $V_0$
in this un-coupled scenario is given by \eq{V3} to be $V_0=(\mu^2K+\nu^2|b|^2)/2$, namely,
\be
(V_0)_I&=&-\({\mu^4\over 4\a}+{\nu^4\over 4\r}\),\nn\\
(V_0)_{II}&=&-\({\mu^4\over 4(\a+\b m^2)}+{\nu^4\over 4\r}\),\quad (m=1,2),\quad {\rm or}\nn\\
(V_0)_{III}&=&-\({\mu^4\over 4(\a+\g)}+{\nu^4\over 4\r}\),\labels{uncoupledV}\ee
depending on which of the energies is the smallest.
If $\b, \g>0$, the lowest energy occurs in phase {I}. If $\b<0$, {II} has a lower energy than { I}, and $m=2$ always has a lower energy
than $m=1$, hence $m=1$ is never the condensate and we will consider it no further. If $\g<0$, then { II} is still the condensate
if $4|\b|>|\g|$, otherwise { III} is the condensate. The phase diagram based on this discussion is shown in Fig.~1.

\begin{figure}[ht] 
  \centering
  \includegraphics[width=5.19in,height=1.69in,keepaspectratio]{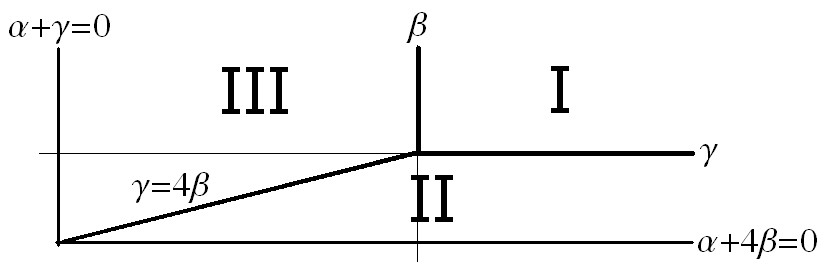}
  \caption{Phase diagram}
  \label{fig:phase}
\end{figure}

If $\t$ is not zero,  the formulas are complicated and the detail is relegated to Appendix B.
Nevertheless, it turns out that Fig.~1 remains valid for all $\t$, at least for those solutions defined
and connected continuously to the un-coupled solutions.

\subsection{Symmetry of Solutions}
Each of the three solutions in Table 2 
is invariant under some symmetry group $\G\subset U(1)\x SO(3)$. 
One way to determine the symmetry is to use Wigner's ${\cal D}$-matrix to compute
$\sum_{m'}a_{m'}{\cal D}^{J=2}_{m'm}(\a',\b',\g')$, and equate it to $a_{m}e^{-i\rho'}$ for some $U(1)$ phase $e^{-i\r'}$. 
The Euler angles of rotations are denoted here by $\a', \b', \g'$, with a prime  to tell them apart  from the coupling
constants in \eq{V} and \eq{abc}.
This gives us five
equations ($-2\le m\le 2$) to solve for four unknowns $\a',\b',\g', \r'$. 
If there is no solution other than $\a'=\b'=\g'=0$, then there is no non-trivial symmetry.
Otherwise, the solution of $(\a',\b',\g')$ yields the $SO(3)$ element and $e^{i\r'}$ the $U(1)$ element in $\G$.

\vskip.5cm

\noindent{\bf I.}\quad  \underline{Tetrahedral Symmetry}
\vskip.3cm
The solution $\f_I=\sqrt{K/3}(1,0,0,\rd,0)$ has a tetrahedral symmetry. It is invariant under the two $U(1)\x SO(3)$ operations 
\be
g_I=e^{-i\b_0 J_y}e^{-i\pi J_z},\quad \tilde g_I=e^{4\pi i/3}e^{-2\pi iJ_z/3},\labels{ab}\ee
with $\sin(\b_0/2)=\sqrt{2/3}$ and $\cos(\b_0/2)=1/\rt$.
Their explicit spin-2 matrix representations are $\tilde g_I=\o^2{\cal D}^{J=2}(0,0,2\pi/3)=\diag(1,\o,\o^2,1,\o)$, and 
\be
g_I={\cal D}^{J=2}(0,\b_0,\pi)={\small{1\over 9}\pmatrix{1&2\rd&2\rs&4\rd&4\cr
2\rd&5&2\rt&-2&-4\rd\cr
2\rs&2\rt&-3&-2\rt&\rs\cr
4\rd&-2&-2\rt&5&-2\rd\cr
4&-4\rd&2\rs&-2\rd&1\cr}}.
\labels{aa}\ee
They can be used to verify the invariance $g_I\f_I=\f_I$ and $\tilde g_I\f_I=\f_I$. 

These two operators satisfy 
\be g_I^2=\tilde g_I^3=(g_I\tilde g_I)^3={\bf 1},\labels{a4pres}\ee
so they generate an $A_4$ group, the invariant group of a tetrahedron.
 Note that the $U(1)$ phase factor in \eq{ab} is not necessary for
$\tilde g_I$ to be a generator of $\G_I=A_4$, but it is necessary for $\f_I$ to be invariant under $\tilde g_I$.

\vskip.5cm
\noindent{\bf II}.\quad \underline{Ferromagnetic Symmetry\ ($m=2$)}
\vskip.3cm
The solution $\f_{II}=\sqrt{K}(1,0,0,0,0)$ of \eq{A11} is invariant under the operation $g_{II}=u(2\xi)e^{-i\xi J_z}$ for any $\xi$,
hence $\G_{II}=SO(2)$.

Originally there were two additive quantum numbers, the horizontal charge $Q$ and the third component of the horizontal spin $m$. 
In the ferromagnetic phase, only $m'=m-2Q$ is conserved.
This is reminiscent of the Standard Model, in which both hypercharge and weak isospin are broken in the Higgs phase, leaving the electric
charge to be the only additively conserved quantum number.

\vskip.5cm

\noindent{\bf III}.\quad \underline{Nematic Symmetry}

The solution 
$\f_{III}=\sqrt{K}(\sin\eta/\rd,0,\cos\eta,0,\sin\eta/\rd)$ 
is invariant under $g_{III}=e^{-\pi i J_z},\ \tilde g_{III}=e^{-\pi i J_y}$, and $\overline g_{III}=e^{-\pi i J_x}$. 
Any two of those three will generate the symmetry group 
$\G_{III}=Z_2\x Z_2$ valid for any $\eta$ in $\f_{III}$.

At special $\eta$'s, the symmetry group is enlarged to $SO(2)$. At $\eta=0$, $\f_{III}\propto(0,0,1,0,0)$ is invariant under any rotation
about the $z$-axis. At $\eta=\pi/3$, $\f_{III}\propto(\rt,0,\rd,0,\rt)$ is invariant under any rotation about the $y$-axis, and at 
$\eta=2\pi/3$, $\f_{III}\propto(\rt,0,-\rd, \rt)$ is invariant under any rotation about the $x$-axis. 

\vskip.5cm
These symmetries of the three phases
 are summarized in Table 3. Note that the generators need a $U(1)$ factor in { I} and { II}, but not { III}.
$$\ba{|c|c|c|c|c|}\hline
{\rm sol}&\eta&\f&\G&g, \tilde g\cr\hline
{\bf I}&&\sqrt{K\over 3}\(1,0,0,\rd,0\)&A_4&e^{-i\b_0 J_y}e^{-i\pi J_z},\ e^{4\pi i/3}e^{-2\pi iJ_z/3}\cr\hline
{\bf II}&&\sqrt{K}(1,0,0,0,0)&SO(2)&u(2\xi)e^{-i\xi J_z}\cr\hline
{\bf III}&\eta&\sqrt{K}\({1\over\rd}\sin\eta,0,\cos\eta,0,{1\over\rd}\sin\eta\)&Z_2\x Z_2&e^{-\pi iJ_z},\ e^{-\pi iJ_y}\cr\cline{2-5}
&0&\sqrt{K}(0,0,1,0,0)&SO(2)&e^{-i\xi J_z}\cr\cline{2-5}
&\pi/3&\sqrt{K\over 8}\(\rt,0,\rd,0,\rt\)&SO(2)&e^{-i\xi J_y}\cr\cline{2-5}
&2\pi/3&\sqrt{K\over 8}\(\rt,0,-\rd,0,\rt\)&SO(2)&e^{-i\xi J_x}\cr
\hline\ea$$
\bc Table 3. Symmetry of the spin-2 solutions\ec

\section{Mass Matrix and Mixing}
We will discuss in this section how to compute mass matrix and fermion mixing for the three phases. Phase I with an $A_4$ symmetry is
presumably suitable for leptons. Phases II and III with a hierarchical structure are potentially good for the quarks, but
phase II turns out to be unphysical because its main mixing is between a heavy and a light quark.

The mass matrix in these phases is most easily computed when  valons are expressed in a matrix
form. The following subsection shows how this can be done for valons of horizontal spins 0, 1, and 2. 

\subsection{Valons in matrix form}
In the Cartesian
basis,
a spin-2 valon is described by a traceless symmetric matrix (or tensor) $\Phi=(\Phi_{ij})\ (i, j=1, 2, 3$ or $x, y, z)$,
whose  entries are given by the components of the spin-2 state $\f=(a_2,a_1,a_0,a_{-1},a_{-2})$ to be
\be
\Phi=\pmatrix{{1\over 2}(a_2+a_{-2})-{1\over\rs}a_0&{i\over 2}(a_{-2}-a_2)&{1\over 2}(a_{-1}-a_1)\cr
{i\over 2}(a_{-2}-a_2)&-{1\over 2}(a_2+a_{-2})-{1\over\rs}a_0&{i\over 2}(a_{-1}+a_1)\cr
{1\over 2}(a_{-1}-a_1)&{i\over 2}(a_{-1}+a_1)&{2\over\rs}a_0\cr 
}.\labels{symtra}\ee
The normalization is chosen so that $K:=\sum_m|a_m|^2=\Tr(\Phi^\dagger\Phi)$. With that normalization,
the scalar $\th$ is given by $\th=\sum_m(-)^ma_ma_{-m}=\Tr(\Phi^2)$. The three components of the vector $\bk{\vec J}$ is proportional to 
the three independent components of the anti-symmetric hermitian matrix $[\Phi^\dagger,\Phi]$,
with $\bk{\vec J^{\ \!2}}=\bk{J_z}^2=(\sum_mm|a_m|^2)^2=-2\Tr([\Phi^\dagger, \Phi]^2)$.

Valons of spin 0 and spin 1 may also be expressed as matrices.
The matrix form $B$ of a spin-0 valon $b$, and $\Gamma$ of a spin-1 valon $c=(c_1,c_2,c_3)$ in the Cartesian basis, are
\be
B={\small{1\over\rt}\pmatrix{b&0&0\cr 0&b&0\cr 0&0&b\cr}},\quad \Gamma={\small {1\over\rd}\pmatrix{0&c_3&-c_2\cr -c_3&0&c_1\cr c_2&-c_1&0\cr}}.
\labels{BG}\ee
Normalization is chosen so that $\Tr(B^\dagger B)=|b|^2$ and $\Tr(\Gamma^\dagger\Gamma)=\sum_i|c_i|^2:=|c|^2$.

Under an $SO(3)$ transformation, $A\to h^\dagger Ah$ for $A=\Phi, B$, and $\Gamma$, where $h$ is the three-dimensional representation of the $SO(3)$ rotation {\it in the Cartesian basis}, obtained from ${\cal D}^{J=1}(\a',\b',\g')$ through a similarity transformation  
that converts spherical basis into Cartesian basis.

In particular, 
the matrix expressions of the three condensates $\f_I, \f_{II}$, and $\f_{III}$ are
\be
\Phi_I={\small {1\over 2}\sqrt{K\over 3}\pmatrix{1&-i&\rd\cr -i&-1&\rd i\cr \rd&\rd i&0\cr}},\quad
\Phi_{II}={\small {\sqrt{K}\over 2}\pmatrix{1&-i&0\cr -i&-1&0\cr 0&0&0\cr}},\nn\\ \nn\\ 
\Phi_{III}={\small\sqrt{K\over 6}\pmatrix{-\cos\eta+\rt\sin\eta&0&0\cr 0&-\cos\eta-\rt\sin\eta&0\cr 0&0&2\cos\eta\cr}}.\labels{Phis}
\ee

For a generic $\eta$, $\Phi_{III}$ is  invariant under the rotations generated by any pair of the  inversions 
$x\to -x,\ y\to -y,\ z\to -z$, reflecting its nematic structure $Z_2\x Z_2$.

For $\Phi_I$ and $\Phi_{II}$, the explicit form of $h_I, \tilde h_I$,
$h_{II}$, the three-dimensional Cartesian representations of the  $SO(3)$ operators
 $g_I=e^{-i\b_0J_y}e^{-\pi i J_z}$, $\o\tilde g_I=e^{-2\pi i J_z/3}$, 
$u(-2\xi)g_{II}=e^{-i\xi J_z}$, are
\be
h_I={\small{1\over 3}\pmatrix{1&0&2\rd\cr 0&-3&0\cr 2\rd&0&-1}},\quad 
\tilde h_I={\small{1\over 2}\pmatrix{-1&-\rt&0\cr \rt&-1&0\cr 0&0&2\cr}},\quad
h_{II}={\small\pmatrix{\cos\xi&-\sin\xi&0\cr \sin\xi&\cos\xi&0\cr 0&0&1}}.
\labels{fff}\ee
Using them, the correct transformations
\be
h_I^\dagger \Phi_I h_I=\Phi_I,\quad \tilde h_I^\dagger\Phi_I\tilde h_I=\o\Phi_I;\qquad
h_{II}^\dagger \Phi_{II}h_{II}=e^{-2i\xi}\Phi_{II}\labels{ftrans}\ee
can be verified to be true.

\subsection{Fermion Mass Matrices}
With the left-handed  and right-handed fermions taken to be horizontal vectors, the mass matrix $M$ 
is just  the expectation values of valons they couple to, expressed
in the matrix form:
\be M=\Phi+B+\Gamma.\labels{mass}\ee
The fields $a_m, b, c_i$ in this formula should be interpreted as expectation values.
 We will assume they
 can be assigned any value that suits the phenomenology.
No Yukawa coupling constant appears because they have been absorbed into the fields.

In order to gain a qualitative understanding of the three phases, let us examine the simplest case when the expectation values are
given only by the condensates in the last section. Namely, $\Phi$ is given by \eq{Phis}, and $B=\Gamma=0$.

The squared-mass spectra in these three phases can then be obtained from the eigenvalues of $\overline M:=M^\dagger M=\Phi^\dagger\Phi$ to be
\bn
\i[{I.}]\quad ${K\over 3}, {K\over 3}, {K\over 3}.$
\i[{II.}]\quad 0, 0, $K$
\i[{III.}] \quad $K(\ol s-\ol c)^2,\ K(\ol s+\ol c)^2,\ 4K\ol c^2$, where $\ol s=\sin\eta/\rd$ and $\ol c=\cos\eta/\rs$.
\en

On the surface solution II seems to be tailored made for quark mixing:
the mass matrix \eq{Phis} has a block diagonal form and the mass spectrum is hierarchical, with one heavy mass $K$ and two light
masses 0. The trouble is, one of the light masses comes from the unmixed 33 entry.
Since both the up- and the down-quark mass matrices have these properties, quark mixing occurs only
in the 1-2 block, giving rise to a mixing between a heavy quark
and a light quark, not between two light quarks.  
The inclusion of
$B$ and $\Gamma$ will not change the behavior qualitatively, so phase II has to be rejected
on phenomenological grounds. However, the remaining
two phases will be put to good use as we shall see in the next two subsections.

\subsection{Quark Sector (Phase III)}
In order to force the mixing to be between the first two generations, we will assume only $c_3\not=0$ when the vector acquires an
expectation value. In that case
the mass matrix is
\be
M=\Phi_{III}+B+\Gamma={\small \pmatrix{\sqrt{K}(-\ol s+\ol c)+b'&c_3'&0\cr -c_3'&-\sqrt{K}(\ol s+\ol c)+b'&0\cr 
0&0&2\sqrt{K}\ol c+b'\cr}},\labels{quark}\ee
where $\ol c=\cos\eta/\rs,\ \ol s=\sin\eta/\rd,\ b'=b/\rt$, and $c_3'=c_3/\rd$.
This matrix has the correct block-diagonal form. It can be made hierarchical by adjusting the three parameters $K, b, \eta$
so that the 33-entry is much bigger than the 11- and 22-entries.
We will assume $\eta$ to be in the second quadrant so that the hierarchical mass spectrum $m_1<m_2<m_3$ can be obtained with
 $b'$ and $\ol s$  positive, $\ol c$ negative, and $|\ol s|\gg|\ol c|$. 
We shall also assume $c_3':=i\g$
to be purely imaginary. In that case $M$ is hermitian and it can be diagonalized by a unitary matrix.

Since the third generation stands by itself, we need to consider only the submatrix $m_{12}$ in the upper-left block. It can
be diagonalized using the formula 
\be{\small
\pmatrix{\cos\th& -i\sin\th\cr -i\sin\th&\cos\th\cr}\pmatrix{\a&i\g\cr -i\g&\b}\pmatrix{\cos\th& i\sin\th\cr i\sin\th&\cos\th\cr}=
\pmatrix{m_1&0\cr 0&m_2\cr}},\labels{diagon}\ee
where
\be
&m_1+m_2=\a+\b=2(b'-\sqrt{K}\ol s),\quad m_2-m_1=\sqrt{(\b-\a)^2+4\g^2}=-2\sqrt{K}\ol c&.\nn\\
 &\tan 2\th={2\g\over\b-\a}&,\labels{mth}\ee
The constants $K, b, \eta$ can be fixed from the three quark masses $m_1, m_2, m_3$. Adjusting the relative size between 
$\g$ and $\b-\a$, the angle $\th$ can be chosen to be anything we want. The diagonalization matrix is unitary but not orthogonal,
but it differs from an orthogonal matrix just by some column and row phase factors which we are allowed to insert by adjusting the
phases of the up and down quarks. Thus the Cabibbo angle is just the difference of the $\th$-angles for the down and the up quarks,
something that is up to us to adjust.

This simple quark model is presented  to illustrate that Cabibbo mixing
with hierarchical quark mass matrices can be accommodated in the nematic 
phase III. 

\subsection{Lepton Sector (Phase I)}
Before plunging into the construction of a horizontal-spin model of leptons in phase I, it is useful to
recall the representations of $A_4$, and how trimaximal as well as tri-bimaximal mixings are obtained.

$A_4$ has four irreducible representations (IR), \ul 1,\
$\ul 1',\ \ul 1'',$ and ${\ul 3}$. The representation of its generators $g_I$ and 
$\tilde g_I$, satisfying the relations $g_I^2=\tilde g_I^3=(g_I\tilde g_I)^3=1$, is given in Table 4 
in the basis where $\tilde g_I$ is diagonal ($\o=e^{2\pi i/3}$).
$$\ba{|c|c|c|c|c|}\hline
{\rm IR}&{\ul 1}&{\ul 1'}&{\ul 1''}&{\ul 3}\cr\hline
\tilde g_I&1&\o&\o^2&\diag(1,\o,\o^2)\cr\hline
g_I&1&1&1&{1\over 3}{\scriptsize\pmatrix{-1&2&2\cr 2&-1&2\cr 2&2&-1}}\cr
\hline\ea$$

\vss
\bc Table 4. Irreducible representations  of  $A_4$\ec

We shall refer to the basis used in Table 4 as the `$F$ basis', and denote the \ul 3-representation of 
$g_I$ and $\tilde g_I$ there as $j_I$ and $\o^2\tilde j_I$. They differ from the corresponding
$SO(3)$ elements $h_I$ and $\tilde h_I$ in \eq{fff} only by a unitary transformation,
\be
u:={\footnotesize{1\over\rd}\pmatrix{1&i&0\cr 1&-i&0\cr 0&0&\rd}},\quad h_I=u^\dagger j_I u,\quad \tilde h_I=u^\dagger\tilde j_Iu.\labels{hju}\ee

When both the left-handed and the right-handed fermions are $A_4$ triplets, the Yukawa valons coupled to them belong to
 $\ul 1,\ \ul 1',\ \ul 1'',\ \ul 3_s,$ or $\ul 3_a$, where $\ul 3_s$ is a triplet that couples symmetrically to the two fermions,
and $\ul 3_a$ a triplet that couples anti-symmetrically. To obtain trimaximal mixing, 
we should assign the expectation value of a valon in the charged-lepton (neutrino) sector
to be an invariant eigenvector of $\tilde g_I\ (g_I)$, for every irreducible representation. 
If that is not possible then the expectation value shoud be taken to be zero. With this assignment trimaximal mixing is guaranteed.
Once a trimaximal mixing is thus obtained, an appropriate adjustment of the Yukawa coupling
constants can promote the mixing into a tri-bimaximal mixing.

To implement this procedure
in the $U(1)\x SO(3)$ theory where $g_I=e^{-i\b_0 J_y}e^{-i\pi J_z}$  and  $\tilde g_I=\o^2e^{-2\pi iJ_z/3}$,
it is necessary to identify the nine $A_4$ states in representations
$\ul 1,\ \ul 1',\ \ul 1'',\ \ul 3_s,\ \ul 3_a$ from the nine horizontal states endowed with horizontal
spins 0, 1, and 2. The spin-0 and spin-1 valons are $B$ and $\Gamma$ in \eq{BG}. As to the spin-2 state, we assume
it either comes from
the dynamical valon $\Phi':=\Phi-\Phi_I$  acquiring a vacuum expectation value at a lower energy,
 or a new low-energy spin-2 valon which does not participate in the high energy dynamics that determines the phase structure. 
To simplify writing in the first scenario, which we will implicitly assume from now on,
the prime will be omitted. Instead, we will replace $\Phi$ in \eq{mass} by $\Phi_I+\Phi$. Equivalently,
in the spherical basis, $\f$ is to be replaced by $\f_I+\f$.

It is obvious from Table 4 and the explicit form of $g_I, \tilde g_I$ that the spin-0 state is a $\ul 1''$
of $A_4$. 
The spin-2 state $\f$ can be decomposed into a mixture of
$\ul 1, \ul 1', \ul 3$ of $A_4$ as follows. 
 We already know that $(1,0,0,\rd,0)$ 
transforms like $\ul 1$. Using \eq{aa} it is easy to verify that $(0,-\rd,0,0,1)$ transforms like $\ul 1'$. After projecting these two out,
what remains transforms like $\ul 3$. In this way the decomposition of a spin-2 state $\f=(a_2,a_1,a_0,a_{-1}, a_{-2})$ is obtained.
\be
\f_1&=&{1\over 3}(a_2+\rd a_{-1})\(1,0,0,\rd,0\):=\rt e\(1,0,0,\rd,0\),\nn\\
\f_{1'}&=&{1\over 3}(a_{-2}-\rd a_{1})\(0,-\rd,0,0,1\):=\rt d\(0,-\rd,0,0,1\),\nn\\
\f_3&=&{1\over 3}\(2a_2-\rd a_{-1},\ a_1+\rd a_{-2},\ 3a_0,\ a_{-1}-\rd a_2,\ 2a_{-2}+\rd a_1\),\nn\\
\f&=&\f_1+\f_{1'}+\f_3.\labels{psi3}\ee
$\f_3$ is a 5-dimensional vector with 3 independent components. It can be projected into the $\ul 3$ space to get
\be
v=(v_1, v_2,v_{3})={1\over\rt}(- a_1-\rd a_{-2},\ \rt a_0,\ -a_{-1}+\rd a_2):=\f\P. \labels{cprimes}\ee
The components of $v$ are chosen to commute with the $SO(3)$ rotations, 
so that $(\f g_I)\P=(\f\P)f_I$ and $(\f\tilde g_I)\P=(\f\P)\o^2\tilde f_I$,
where 
$f_I$ and $\tilde f_I$ are the {\it $SO(3)$ part} of the spin-1 $A_4$ generators in the spherical base, namely,
$\tilde f_I=\diag(\o^{-1}, 1, \o)$ and
\be f_I=-{\footnotesize{1\over 3}\pmatrix{1&2&2\cr 2&1&-2\cr 2&-2&1\cr}}.\labels{fI}\ee
These two differ from the $\ul 3$ representation $j_I=g_I$ and $\tilde j_I=\o^{-2}\tilde g_I=\diag(\o,\o^2,1)$  in the last column 
of Table 4 only by an 
inconsequential unitary transformation $W$,
\be
W:={\footnotesize\pmatrix{0&0&1\cr -1&0&0\cr 0&1&0\cr}},\quad j_I=W f_I W^\dagger,\quad \tilde j_I=W \tilde f_IW^\dagger.\labels{Wjf}\ee
They also differ from $h_I$ and $\tilde h_I$ of \eq{fff} just by a unitary transformation $C$ that converts
spherical harmonics into Cartesian coordinates:
\be
C=uW^\dagger={\footnotesize{1\over\rd}\pmatrix{-1&i&0\cr 0&0&\rd\cr 1&i&0}},\quad h_I=C^\dagger f_1C,\quad \tilde h_I=C^\dagger\tilde f_1C.\labels{cfh}\ee

The normalization of $v$ is chosen so
that its norm $|v|^2:=\sum_i|v_i|^2=|\phi_3|^2:=\sum_m|(\f_3)_m|^2$.

Besides the spin-2 valon $\f$, the mass matrix $M$ also receives contribution from the spin-0 valon $b$ which is a $\ul 1''$,
and the spin-1 valon $c$ which is a $\ul 3_a$ because $\Gamma$ is an anti-symmetric matrix.

Next, let us compute the expectation value of valons needed to yield trimaximal mixing.
Those used in the charged-lepton mass matrix $M_e$ are given by the invariant eigenvectors of $\tilde g_I$, and those used in the neutrino
mass matrices $M_\nu$ are given by the invariant eigenvectors of $g_I$. In case the operator does not have an eigenvalue $+1$,
the corresponding expectation value is zero. The solution to this requirement is summarized in Table 5.
Normalization cannot be determined so those listed in Table 5 are really vacuum alignments,
computed in the $F$-basis where $\tilde g_I$ is diagonal. 
The vectors $v_F$ and $c_F$ are vectors $v$ and $c$ expressed in that basis. The table also gives
relations between $a_m$ and between
 $c_i$ needed to obtain such vacuum alignments. They can be computed either from the invariant
eigenvectors $v_F$ and $c_F$, or more directly from the condition that $\tilde h_1^\dagger M_e \tilde h_1=\o M_e$ and 
$h_1^\dagger M_{\nu} h_1=M_{\nu}$.
$$\ba{|c|c|c|c|c|c|}\hline
{\rm valon}&e&d&v_F&b&c_F\cr\hline
{\rm IR}&\ul 1&\ul 1'&\ul 3&\ul 1''&\ul 3\cr\hline\hline
M_e&1&0&(1,0,0)&0&(1,0,0)\cr\cline{2-6}
&&&a_1=a_0=a_{-2}=0&&c_1=c_3=0\cr\hline\hline
M_\nu&1&1&(1,1,1)&1&(1,1,1)\cr\cline{2-6}
&&&-\rd a_2+a_1+a_{-1}+\rd a_{-2}=0&&c_1=\rd c_3\cr
&&&-\rd a_2+\rt a_0+a_{-1}=0&&c_2=0\cr
\hline\ea$$
\bc Table 5. Vacuum alignment of horizontal valons\ec

Using Table 5, we obtain the $F$-basis mass matrix $u(\Phi_I+\Phi+B+\Gamma)u^\dagger$ to be
\be
M_e&=&{\footnotesize\pmatrix{0&a'_2&0\cr 0&0&a'_{-1}/\rd-\rd c_2\cr a'_{-1}/\rd+\rd c_2&0&0\cr}},\nn\\ \nn\\
M_\nu&=&{\footnotesize\pmatrix{-v_3'+b'-ic_3&e'+2v_3'&d-v_3'+ic_3\cr  d+2v_3'&b'-v_3'+ic_3&e'-v_3'-ic_3\cr e'-v_3'+ic_3&d-v_3'-ic_3&b'+2v_3'\cr}},
\labels{menu}\ee
where $\kappa=\sqrt{K/3},\ a_2'=a_2+\kappa,\ a_{-1}'=a_{-1}+\rd\kappa,\ e'=e+\kappa,\ b'=b/\rt$,
 and $v_3'=v_3/\rs$. 

In this $F$-basis, $j_IM_\nu j_I^\dagger=M_\nu$ 
and $\tilde j_I^\dagger M_e \tilde j_I=\o M_e$. Although $\tilde j_I$ is diagonal,
$M_e$ is not because of the phase factor $\o$. However, this phase factor cancels out in  $\ol M_e=M_e^\dagger M_e$
so $\ol M_e$ is diagonal, with eigenvalues
$m_e^2=(\rd c_2+a_{-1}/\rd)^2,\ m_\mu^2=a_2^2,\ m_\tau^2=(\rd c_2-a_{-1}/\rd)^2$.
They can be used to determine the parameters $c_2, a_2, a_{-1}$ from the masses of the charged leptons.  
In particular, $c_2$ and $a_{-1}$ must have opposite signs to yield
the correct hierarchy of masses.

The expression for $M_\nu$ in \eq{menu} can be used for the Dirac mass matrix of the neutrinos as well as the Majorana mass matrix
of the heavy right-handed neutrinos, with two different sets of parameters of course. This matrix is {\it magic} because the sum of every
row and every column is equal to $b'+d+e'$, hence the mixing matrix is trimaximal \cite{23magic}. If we choose the parameters such
that $e'=d$ and $c_3=-3iv_3'$, then the mixing matrix is 2-3 symmetric and hence bimaximal \cite{23magic}. With this choice, the 
neutrino mixing
matrix is both trimaximal and bimaximal so it has the TBM form.

\section{Conclusion}
We propose to resort to dynamics to determine which of the finite non-abelian subgroups of $SO(3)$ is the appropriate horizontal symmetry
 of leptons. We do so by using dynamics of valons of horizontal spin not larger than 2, invariant under
the smallest group containing $SO(3)$ that can spontaneously break down to at least one non-abelian subgroup. It is found that
using $U(1)\x SO(3)$, the only non-abelian subgroup it can break down to  is $A_4$, thereby making it the
preferred horizontal symmetry of leptons. Other than  $A_4$, this dynamics also produces two other phases, 
a phase II with a $SO(2)$ symmetry,
and a phase III with a $Z_2\x Z_2$ symmetry, both capable of  describing Cabibbo mixing of the quarks with a mass hierarchy.
 It turns out that phase II yields the wrong mass hierarchy and has to be rejected, but phase III is perfectly viable. 
Thus $U(1)\x SO(3)$ can also be thought of as a common high-energy horizontal symmetry where both the Cabibbo mixing of quarks and 
the tri-bimaximal
mixing of neutrinos originate.
Explicit mass matrices
in the quark sector (phase III) and in the lepton sector (phase I) are constructed to illustrate these features. 

Technical complication prevented us from including a fully coupled spin-1 valon into the potential in this paper.
CP violation is also left out. To include the latter it is presumably necessary to go beyond $SO(3)$ to $SU(3)$. These points are
being investigated. There are also several other dynamical issues that requires further study. For example, we have assigned phase I
to the lepton sector and phase III to the quark sector because that seems to be what phenomenology demands, but it would be
much nicer if a dynamical mechanism can be found to force that to happen automatically. 
This paper is of an exploratory nature, 
to test the idea and to demonstrate
the feasibility of a symmetry group common to the quark and lepton sectors, a higher group that can 
even select the lepton symmetry at lower energies. It is not meant to present a complete dynamical model so detailed questions 
such as whether this higher symmetry is local or global,
and the associated question of Goldstone bosons, etc., have all been left out.

I am grateful to Fei Zhou, We-Fu Chang, and James Bjorken for stimulating discussions. 
Part of this work was carried out while visiting the Theoretical Science 
Center at Tsinghua University in Hsinchu, and the Academia Sinica in Taipei, in the summer of 2010, whose hospitality I would also like
to acknowledge.

\newpage
\appendix
\section{Uniqueness of the spin-2 solutions}
The purpose of this appendix is to show that the $\bk{\vec J^{\ \!2}}=\bk{J_z}^2=0$ solutions \eq{psi0} and \eq{nematic} are unique up to $U(1)\x SO(3)$ transformations.

As discussed in Sec.~IV, a spin-2 valon $\f$ can be represented by a traceless symmetric matrix $\Phi$, with 
$\bk{\f|\f}=\Tr(\Phi^\dagger\Phi)$, $\th=\Tr(\Phi^2)$, and $\bk{\vec J^{\ \!2}}=-2\Tr([\Phi^\dagger,\Phi]^2)$.

Since $[\Phi^\dagger,\Phi]$ is hermitian, $[\Phi^\dagger,\Phi]^2$ is positive semi-definite,
with non-negative real eigenvalues. To have a zero trace needed for $\bk{\vec J^{\ \!2}}=0$,
 all eigenvalue must be zero, implying $[\Phi^\dagger,\Phi]=0$.
Therefore $\Phi^\dagger$ and $\Phi$ can be simultaneously diagonalized by an orthogonal transformation.

Suppose $\diag(a,b,-a-b)$ is the diagonal form of the traceless matrix $\Phi$. Applying a $U(1)$ transformation 
if necessary we may assume $a$ to be real. The $\th$ parameter is given by
\be
\th=\Tr(\Phi^2)=2(a^2+b^2+ab).\labels{appth}\ee

If $\th=0$, then it is still zero if we multiply both sides of \eq{appth} by $a$ or by $b$. Equating these two identities, we get
$a^3=-a(b^2+ab)=-b(ab+a^2)=b^3$. Therefore $b/a$ is a third root of unity. This root cannot be 1 for then $\th\not=0$, hence
it must be $\o=e^{2\pi i/3}$ or $\o^2=e^{4\pi i/3}$. Thus the eigenvalues of $\Phi$ are $a,\ a\o,$ and $a\o^2$.
Since $a$ is fixed by the norm $K=\bk{\f|\f}=\Tr(\Phi^\dagger\Phi)$ to be $a=\pm\sqrt{K/3}$, the solution $\Phi$ for $\th={J_z}=0$ is unique
up to a sign and a $U(1)\x SO(3)$ transformation, hence $\f_I$ in \eq{psi0} is unique up to a $U(1)\x SO(3)$ transformation.

As a check, we can compute directly the eigenvalues of $\Phi_I$.
They turn out to be $\sqrt{K/3},\ \sqrt{K/3}\ \o$, and $\sqrt{K/3}\ \o^2$.

If $|\th|=K$, with a $U(1)$ transformation we can render $\th=K$. In that case $\th=a^2+b^2+(a+b)^2=K=|a|^2+|b|^2+|a+b|^2$.
Since $a$ is real, so must be $b$. As a function of $b$, $a$ reaches a stationary point at $a=-2b$, whence $\th=3b^2=K$, giving
$b=\pm\sqrt{K/3}$ and $a=\mp 2\sqrt{K/3}$. We can now parametrize $a$ to be
$a=2\sqrt{K/3}\cos\eta$. Substituting this into \eq{appth} with $\th=K$, we can solve
$b$ to get $b=\sqrt{K/3}(-\cos\eta\pm \rt\sin\eta)$. Thus the eigenvalues of $\Phi$ are $\sqrt{K/3}(2\cos\eta, -cos\eta+\rt\sin\eta,
-\cos\eta-\rt\sin\eta)$. Comparing this with the expression for $\Phi_{III}$ in \eq{nematic}, we see that the $\eta$ here is the same
as the $\eta$ there. It also shows the uniqueness of $\q_{III}$ up to a $U(1)\x SO(3)$ transformation.

\section{Interacting condensates}
The dependence of $V$, $K,\ |b|^2$ on  $\t$ and $\s$  will be
 worked out in this appendix for phases I, II, and III. 
We will also work
out what the phase diagram Fig.~1 becomes when $\tau\not=0$.

For the sake of this discussion,
we shall assume
the spin-2 condensate in the absence of coupling
 to be larger than the spin-0 condensate, namely, $\sqrt{\a}\nu^2/\sqrt{\r}\mu^2:=r<1$.

\subsection{$\th=0,\ \bk{J_z}=0$}
The condensates are given in \eq{A2} and the energy in \eq{A3}. 

First consider how $V_I$ changes with $\t$. At $\t=0$, 
$V_I=(V_0)_I=-(\mu^4/4\a+\nu^4/4\r)$. Since $(dV_I/d\t)_{\t=0}=\mu^2\nu^2/4\a\r$ is positive,
$|V_I|$ decreases with increasing $\t$ in the vicinity of $\t=0$. Farther away, $V_I$ has two
stationary points where $dV_I/d\t$ vanishes. The smaller one is located at $\t=\t_-:=2\a\nu^2/\mu^2:=\t_0r$, where $\t_0:=2\sqrt{\a\r}$,
and the larger one is located at $\t_+=\t_0/r$. $V_I$ reaches
a local maximum $(V_I)_-=-\mu^4/4\a$ at $\t_-$ and a local minimum $(V_I)_+=-\nu^4/4\r$ at $\t_+$. The local maximum is smaller
(lower) than the local minimum because $r<1$. The sum of the values at the local maximum and local minimum happens to be the value $V_I$
at $\t=0$.

There are two asymptotes located at $\t=\pm \t_0$ where $V_I=\pm\infty$. They divide the graph of $V_I$ vs $\t$ into three
branches. Since $0<\t_-=\t_0r<\t_0$,
the smaller stationary point appears in the central branch. The other stationary point at
$\t_+=\t_0/r>\t_0$ is to be found in the right-hand branch.

The left-hand branch starts at $V_I=0$ at $\t=-\infty$ and rises monotonically to $+\infty$ at the left asymptote $\t=-\t_0$.
The central branch starts at $V_I=-\infty$ at the left asymptote, rises to a maximum at $\t=\t_-$, then drops back to $-\infty$ at the right
asymptote. The right-hand branch starts at $V_I=+\infty$ at the right asymptote, descends to a minimum at $\t=\t_+$, then rises back to 0
at $\t=+\infty$. 

The central branch is sketched in Fig.~2.

\begin{figure}[h] 
  \centering
  \includegraphics[width=4in,height=3.12in,keepaspectratio]{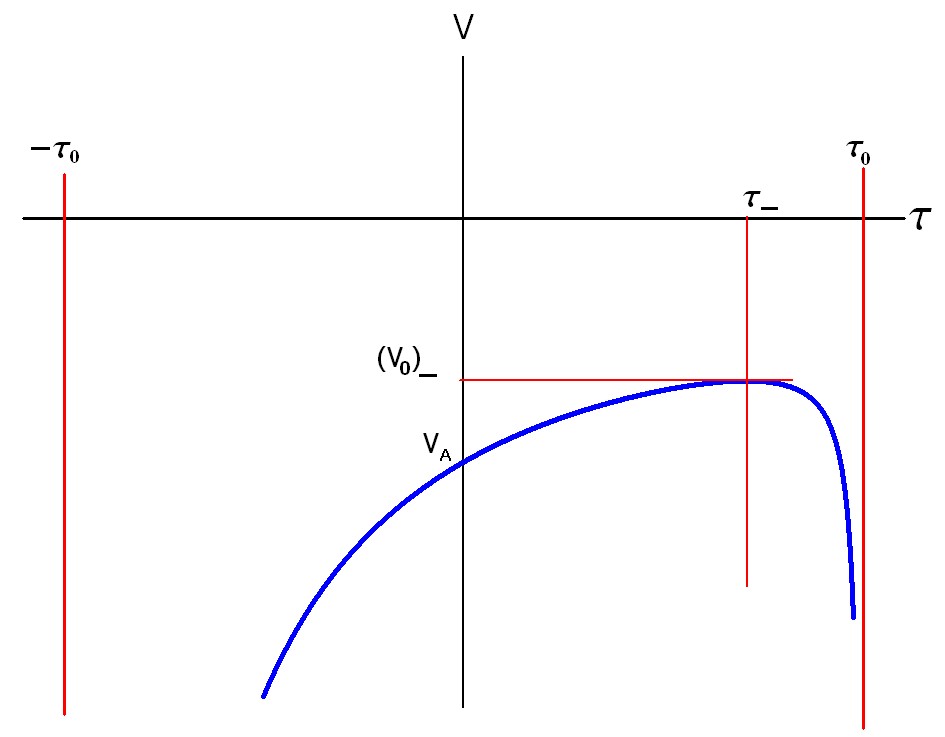}
  \caption{The center branch of the Higgs potential $V_I$ vs $\t$}
  \label{fig:V2}
\end{figure}

Next, let us see how the condensates $b^2$ and $K$ vary with $\t$. At $\t=0$, $b^2=\nu^2/2\r$ and $K=\mu^2/2\a$. 
The product of these two happens to be the slope of $V_I$ at that point.
Both of the $b^2$- and $K$-slopes at
$\t=0$ are negative. $b^2$ decreases monotonically from the value $\nu^2/2\r$ at $\t=0$ to the value 0 at $\t=\t_-$, beyond which 
$b^2$ becomes negative
and unacceptable. For that reason we need not consider the central branch in Fig.~2 to the right of $\t_-$. The condensate $K$
starts from the value $\mu^2/2\a$ at $\tau=0$, decreases to a minimum value $(\mu^2/2\a)(r/2\zeta)$  at
 $\t=\t_m=\zeta\t_0$, where $\zeta:=(1-\sqrt{1-r^2})/r$. 
From $\t=\t_m$ to $\t_-$, $K$ increases from this minimum back to the value $\mu^2/2\a$.
To be consistent, we should have $\zeta<1$ and  $r<2\zeta$, which indeed
follows from the condition $0<r<1$.

For negative $\t$, both $b^2$ and $K$ increase monotonically to the value $+\infty$ at the left asymptote $\t=-\t_0$.

The behavior of $b^2_I,\ K_I$, and $V_I$ is summarized in Table 5. The definition of various $\t_a$ is given
in the second row, with $r:=\sqrt{\a}\nu^2/\sqrt{\r}\mu^2$ and $\zeta:=(1-\sqrt{1-r^2})/r$.  Also,
$\xi:=r^2/2+r/2\zeta-1$. It follows from $0<r<1$ that $0<r/2<\zeta<r$ and $0<\xi<r^2$. Hence $b^2$ and $V$
are monotonic in the range $-\t_0<\t<\t_-$, but $K$ has a minimum located at $\t=\t_m$.

$$\ba{|c|c|c|c|c|}\hline
\tau&-\t_0&0&\t_m&\t_-\\ \hline
{\rm def}&-2\sqrt{\a\r}&0&\zeta\t_0&r\t_0\cr\hline\hline
b^2&+\infty&\nu^2/2\r&\nu^2/4\r&0\cr\hline
K&+\infty&\mu^2/2\a&(\mu^2/2\a)(r/2\zeta)&\mu^2/2\a\cr\hline
V&-\infty&-(\mu^4/4\a)(1+r^2)&-(\mu^2/4\a)(1+\xi)&-\mu^4/4\a\cr
\hline\ea$$
\bc Table 5. The variation of \eq{A2} and \eq{A3} with $\t$ in the case of $\th=\bk{J_z}=0$\ec

\subsection{$\th=0,\ \bk{J_z}=2K$}
The condensates in this phase are given in \eq{A5} and the energy given in \eq{A6}, both with $m=2$. In particular,
$V_{II}$ in \eq{A6}  is the same as $V_I$ in \eq{A3} after having
$\a$ replaced by $\a'=\a+4\b$. 

Since
\be
{\p V_I\over\p \a}={(-2\r\mu^2+\nu^2\t)^2\over(-\t^2+4\a\r)^2}\ge 0,\labels{dvda}\ee
phase II has a higher (lower) energy than phase I if $\b>0\ (<0)$.
In other words,  the boundary between phases I and II in the
phase diagram of Fig.~1 remains the same even if the coupling $\t\not=0$ is turned on.

\subsection{$|\th|=K,\ \bk{J_z}=0$}
The condensates in this phase are given in \eq{A9} and the energy given in \eq{A10}.

If $\s=0$, then \eq{A9} can be obtained from \eq{A3} by replacing $\a$ with 
$\a''=\a+\gamma$. On account of \eq{dvda}, the boundaries of phase III with phases I and II
remain the same as Fig.~1 even in the presence of $\t$-coupling.

If $\s\not=0$, the $\t$-dependence is very complicated. However, if the spin-0 condensate
$b^2$ is non-zero, the solutions in phases I and II do not exist for $\s\not=0$, so
this phase III remains the only phase no matter what $\t$ is.


\end{document}